# Multi-lipid synergy in biolubrication

Yifeng Cao[1,2,†,*], Di Jin[1,†,*], Nir Kampf[1], and Jacob Klein[1*]

1. Department of Molecular Chemistry and Materials Science, Weizmann Institute of Science, Rehovot 76100, Israel

2. Present address: Institute of Zhejiang University-Quzhou, Quzhou 324000, China; College of Chemical and Biological Engineering, Zhejiang University, Hangzhou 310027, China.

\* caoyf@zju.edu.cn; di.jin@weizmann.ac.il; jacob.klein@weizmann.ac.il

† Equal contribution

Abstract:

The very low sliding friction of articular cartilage in the major synovial joints such as hips and knees is crucial for their well-being, and has been attributed to lubrication by phospholipid boundary layers. While single-component lipid layers have demonstrated efficient lubricity in model studies, in living joints there is a large number of different lipids, raising the question of whether this is natural redundancy, or whether this multiplicity confers any benefits. Here we examine lubrication by progressively more complex mixtures of lipids representative of those in joints, using a surface forces balance at physiologically-relevant salt concentrations and pressures. We find that different lipid combinations differ very significantly in their lubricating ability, as manifested by their robustness to hemifusion under physiological loads, pointing to a clear lubrication synergy arising from multiple lipid types in the lubricating layers. Insight into the origins of this synergy is provided by molecular dynamics (MD) simulations of the different lipid mixtures used in the experiments, which directly reveal how hemifusion - associated with greatly increased friction - depends on the detailed lipid composition. Our results provide insight into the role of lipid type proliferation in healthy synovial joints, and point to new treatment modalities for osteoarthritis.

**Introduction**

Synovial joints, such as hips and knees, have evolved as long-lived, highly efficient lubrication systems. Friction at the articular cartilage surfaces coating the joints, which slide past each other as they articulate, is associated with sliding friction coefficients $μ$ ( = [force to slide]/[load]) as low or lower than 0.001[1]. Such low friction is crucial for their homeostasis[1,2], in particular for suppressing wear-related cartilage degradation[3], a leading symptom of osteoarthritis, the most widespread joint pathology. This lubricity has been attributed both to the role of interstitial fluid pressure within the cartilage, and largely to strongly-lubricating boundary layers at the cartilage surface, whose molecular origins have been extensively studied[1,4-7]. In recent years, surface-attached phospholipid layers, in particular phosphatidylcholines (PCs), which are ubiquitous in joints, have been shown to act as exceptionally good boundary lubricants at both synthetic and at biological surfaces[2,8-11]. Model studies have shown that bilayers of zwitterionic PCs[8,11] and sphingomyelin (SM) lipids[12] result in low friction up to physiological pressures, with $μ$ down to $10^{-4}$ or lower, comparable to that of articular cartilage in joints. This arises through the hydration lubrication mechanism[13] active at the highly-hydrated phosphocholine groups exposed by the surface-attached lipid bilayers[14,15] at the slip-plane, and it has been proposed that similar boundary layers are present at the articular cartilage surface, providing its excellent lubricity[1,2,4,16]. To date, studies of boundary lubrication by lipids have all, with few exceptions[17-19], examined single-component PCs[5,8,11,14,20,21]; healthy joints, however, are known to include lipids belonging to many different classes, comprising additionally different tail saturation levels and lengths[22-25].

While lipids have many biological roles, our main interest here is their contribution to joint lubrication. A key question, therefore, is whether - separately from any other functions - this proliferation of different lipids in joints holds benefits for boundary lubrication at the articular cartilage surface, relative to the single or two-component PC layers which have been studied to date. The major phospholipid (PL) groups identified in synovial joints, present in the form of multilamellar and vesicular structure[26,27] include electroneutral PC, SM, and phosphatidylethanolamine (PE), as well as, as minor components, negatively charged PLs[22,24]. The majority have at least one unsaturated tail[24], while the species and concentration of PLs in synovial joints are also affected by joint diseases, such as osteoarthritis (OA) and rheumatoid arthritis[28,29]. In a previous study, we examined the interactions between membranes of a binary mixture of 2 PC lipids: 1,2-dipalmitoyl-*sn*-glycero-3-phosphocholine (DPPC) and 1-palmitoyl-2-oleoyl-glycero-3-phosphocholine (POPC)[19]. In that system hemifusion of the opposing layers, and the consequent elimination of the hydration-lubricated slip-plane at the headgroup-headgroup interface, caused an abrupt increase in the friction force; the hemifusion was attributed to phase separation of gel-phase and liquid phase domains.

The present study focuses on the forces acting between surface boundary layers of lipid mixtures, including the major PL classes found in healthy synovial fluid and on cartilage surfaces, to provide insight into the lubrication efficiency and possible synergy of multi-component lipid boundary layers. By synergy here is meant that the effect of combining different lipids leads to better lubrication than just the sum of the parts. In particular, while it is clearly not possible to test all possible combinations of the order of 100 or more different PLs in joints[22,24,25], a demonstration of improved lubrication arising from particular combinations of different lipids in the boundary layers may provide insight from a biolubrication perspective as to why there is such proliferation of different lipids in joints. Extending our previous work on single-component PLs and binary PC mixtures, we take a first step to mimic the diversity of PLs in joints by adsorbing increasingly complex lipid mixtures (Table 1, and fig. S1 and Table S1) on a model substrate, and measuring normal and frictional forces between them at physiological salt concentrations. Forces are measured using the surface force balance (SFB), where the model substrate is atomically-smooth mica (negatively charged in aqueous media). Detailed molecular dynamics (MD) simulations are then used to shed strong light on the origins of the different lubrication behaviour corresponding to the different lipid combinations.

**Results**

The three PL mixtures composed of 2, 5, and 8 lipids, designated L2, L5, and L8 respectively (Table 1), were selected to include progressively more of the major lipid classes in synovial joints (Fig. S1 and Table S1). To approximately mimic the PLs in synovial joints, all the representative mixtures contain lipids with phosphocholine and phosphoethanolamine headgroups, which are the main lipid groups detected in the joints (as summarized in Fig. S2), and all have unsaturated PLs as majority components. L8 included, in addition to main PL types, also cholesterol (CHOL) which is ubiquitous in physiological lipid bilayers and in synovial fluid.

**Table 1.** Compositions, averaged size, polydispersity (PdI), and zeta potential values of mixed-lipid-SUVs prepared in 150 mM $NaNO_3$ before and after adding $Ca(NO_3)_2$.

| Lipid composition | Dispersant | Size/nm | PdI | Zeta potential/mV |
|---|---|---|---|---|
| **L2:** POPC-POPE (in ratio 4:1) | 150 mM $NaNO_3$ | 68.2 ± 1.2 | 0.064 | -4.8 ± 1.6 |
| | + 2 mM $Ca(NO_3)_2$ | 69.6 ± 2.0 | 0.082 | -2.3 ± 1.7 |
| **L5:** DPPC-POPC-DOPC-POPE-Egg SM (1:1:1:1:1) | 150 mM $NaNO_3$ | 65.1 ± 0.7 | 0.072 | -1.6 ± 0.6 |
| | + 2 mM $Ca(NO_3)_2$ | 64.1 ± 1.1 | 0.066 | -0.4 ± 0.2 |
| | 150 mM $NaNO_3$ | 63.4 ± 0.7 | 0.057 | -7.3 ± 0.3 |
| **L8:** DPPC-POPC-DOPC-POPE-Egg SM-DPPA-LPC-CHOL (1:1:1:1:1:0.1:1:1) | + 2 mM Ca(NO$_3$)$_2$ | Multiple peaks | 0.440 | -5.3 ± 0.3 |

Surface morphologies

To obtain additional insight into the effect of combining the different lipids, all measurements were carried out both in the absence and in the presence of calcium ions (as the Ca(NO$_3$)$_2$ salt) at concentrations (2 mM) similar to their physiological concentrations in healthy synovial fluid. Size distribution and zeta potentials of SUVs of the three mixtures were characterized by DLS and are shown in Table 1, with all three having rather small negative potentials. We point out that addition of 2 mM Ca(NO$_3$)$_2$ had little effect on the DLS-measured vesicle diameters (which remained at ca. 65 nm) for L2 and L5, indicating little aggregation by the divalent ions. In contrast, L8, which included the negatively-charged DPPA (at low concentration – ca. 1.4 mole %) and had a slightly more negative zeta potential, was clearly aggregated by the calcium ions, which presumably acted as adhesive linkers between the DPPA headgroups on the vesicles. Formation of surface assemblies of the lipids on the mica substrate was achieved by spontaneous adsorption of the vesicles from the respective dispersions (Experimental Section, SI), likely driven by the dipole-charge interactions of the majority-components PC-headgroup zwitterions with the mica, following which they ruptured to form bilayers on the mica. Morphologies of these bilayers were characterized by AFM (Fig. 1).

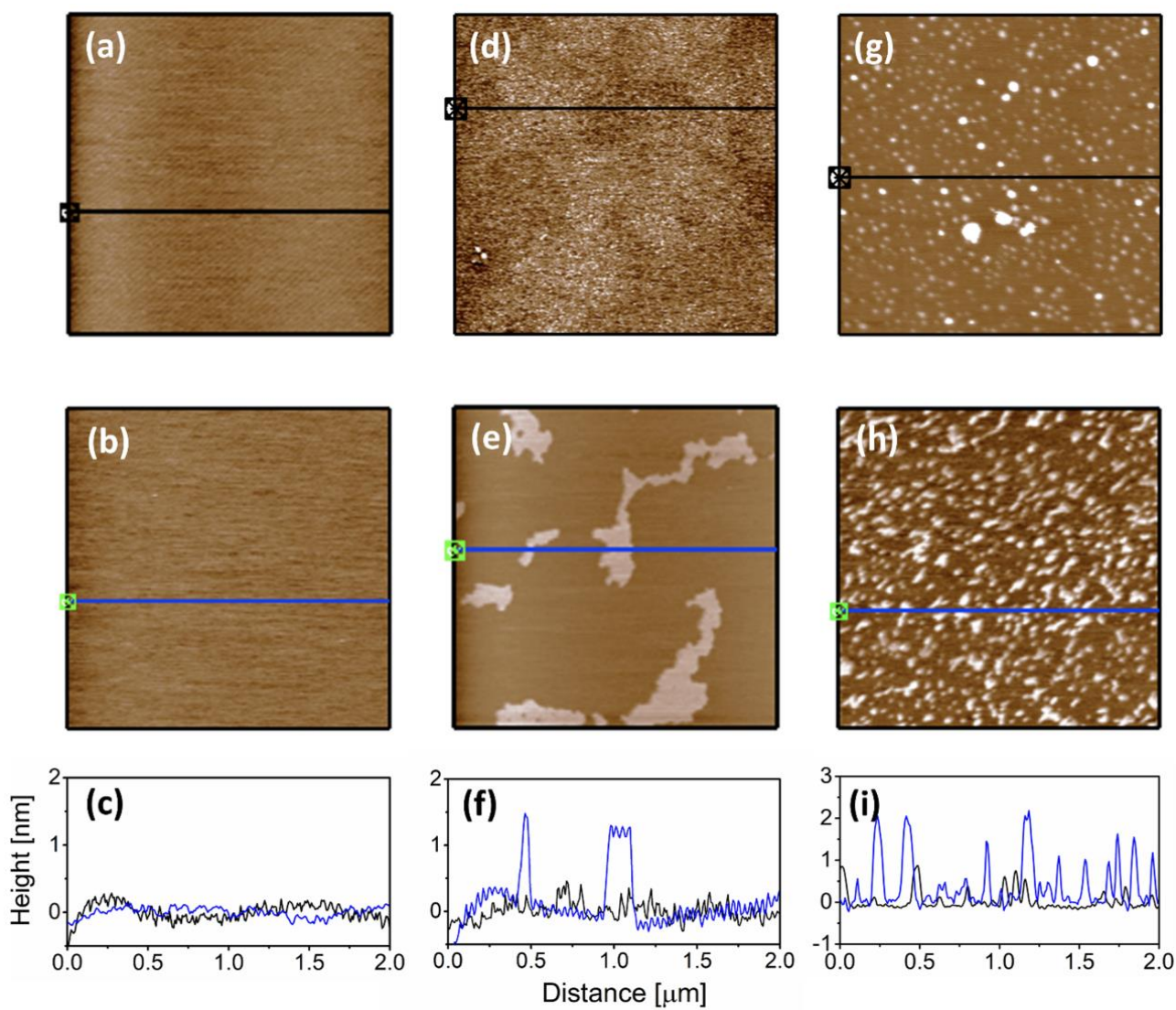

**Figure 1.** AFM height images of SUVs adsorbed on mica before (a, d, and g) and after (b, e, and h) adding 2 mM $Ca(NO_3)_2$ for the L2 (a-c), L5 (d-f), and L8 (g-i) systems, respectively. All scans were performed across 0.3 mM SUV dispersion prepared in 150 mM $NaNO_3$. The size of each image is 2 µm × 2 µm. Panels (c), (f), and (i) in the last row are corresponding height profiles, wherein the black and blue lines represent those before and after adding 2 mM $Ca(NO_3)_2$. The phase-separated (lighter) regions in panel (e) are likely of the gel-phase (saturated) lipid components, DPPC and Egg SM, and occupy ca. 13% of the area scanned as analysed by image segmentation.

Surface interactions

Normal and shear forces between the bilayer-bearing mica surfaces at different surface separations $D$ across the respective SUV dispersions with and without added $Ca(NO_3)_2$ were directly measured using an SFB (fig. 2(a)). Fig. 2 shows typical shear trace profiles for all three mixtures, while Fig. 3 shows their normal and friction force profiles. As seen in figs. 2 and 3, all three mixtures showed broadly similar behaviour both in their normal and in frictional interactions, as follows. The normal interactions $F_n(D)/R$ revealed a short-ranged interaction, as expected at the high salt concentration (150 mM $NaNO_3$) of the dispersions, where the Debye screening length for electrostatic interactions is less than 1 nm. Monotonic repulsions in these profiles commenced at somewhat longer range (ca. 20 – 30 nm for L2, L5 and ca. 50 nm for L8), and may be attributed to steric interactions between loosely adsorbed liposomes on the surface-attached bilayers. At higher loads – emphasized in the insets to figs 3(a), (c), (e) -  the interactions showed 'hard-wall' repulsions at $D \approx 10\pm1$ nm, indicating a bilayer on each surface. As the loads were increased, hemifusion occurred as revealed by a jump-in of the surfaces at a critical normal load, to $D$ values indicating a single bilayer between the surfaces. On adding the calcium salt (to 2 mM $Ca^{++}$ concentration, similar to physiological values), the normal force profiles remained essentially unchanged, save that the critical load required for hemifusion greatly increased. On separation of the surfaces and re-approach, the behaviour was reproducible, indicating healing of any change in the surface morphology due to the hemifusion.

The frictional interactions were also broadly similar for all three mixtures, as seen in fig. 2 and figs. 3(b), (d), (f): At lower loads, all three mixtures had very low friction coefficients μ, down to μ $\approx 10^{-4}$ or lower; when hemifusion occurred as the critical loads were reached, the friction force (and friction coefficient) increased abruptly, as expected. Once $Ca^{++}$ was added, and critical loads for hemifusion became much larger (as detailed below), the friction force upon hemifusion exceeded the maximal applied shear force in the SFB and the surfaces remained in rigid contact when sheared (solid red arrows in figs. 3(b), (d), (f)). In addition to these features of the interaction common to all three mixtures, below we describe additional features specific to each mixture.

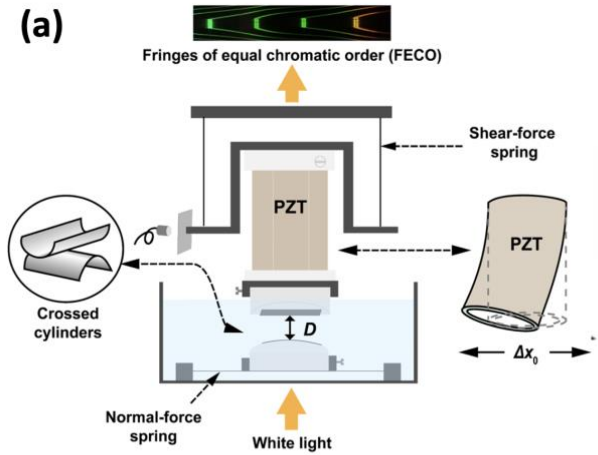

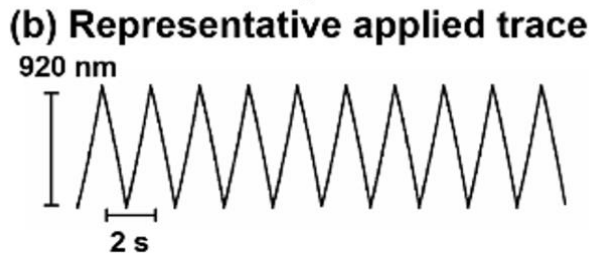

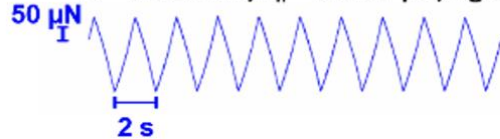

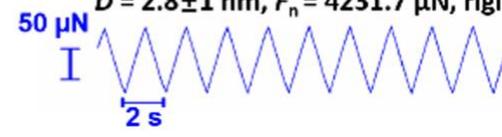

**Figure 2**. Illustration of the SFB setup (a) and representative applied lateral motion to the upper surface (b), and shear force $F_s(t)$ traces (c-e) across L2, L5 and L8 dispersions, respectively. In panel (a), two mica surfaces are in a crossed-cylindrical configuration at a closest separation distance $D$, which is determined according to the wavelengths of fringes of equal chromatic order (FECO) using

the multiple beam interference technique (the fringes shown in the top panel are for two surfaces in adhesive contact). Normal and shear forces were determined by the bending of corresponding springs. Displacement of shear springs is monitored by an air-gap capacitor, as responses to the applied back-and-forth motion via the sectored PZT. In figures (c-e), black and blue traces represent those before and after adding 2 mM Ca(NO$_3$)$_2$. The changes in shear traces at hemifusion, indicated by an increase in $F_s$, were recorded for the L2 system after adding calcium (the lowest blue trace in (c), and for the L5 system before adding calcium (the lowest black trace in (d)). Hemifusion was observed after applying sufficient normal force to the system, and occasionally took place while recording the shear trace, as indicated by the arrows.

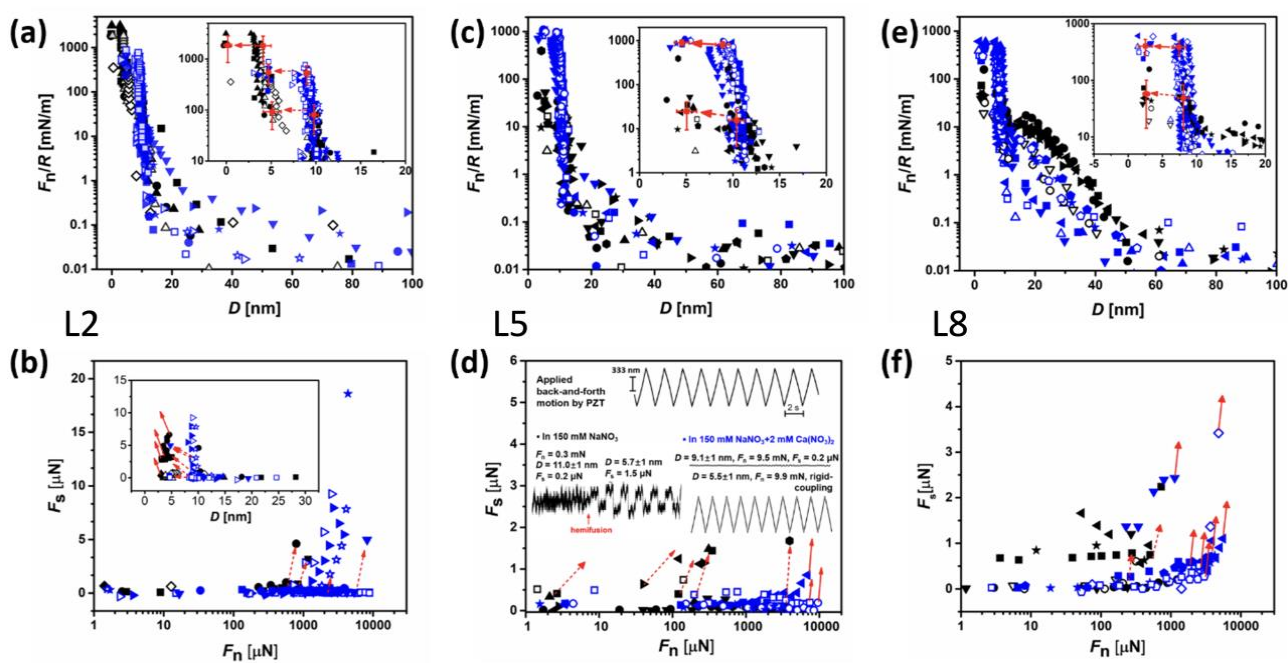

**Figure 3.** Force profiles across 0.3 mM L2 (a and b), L5 (c and d), and L8 (e and f) dispersion in 150 mM NaNO$_3$ before (black symbols) and after (blue symbols) adding 2 mM Ca(NO$_3$)$_2$, respectively. Panels **(a), (c),** and **(e):** Normalized force versus separation distance ($F_n/R$ vs. $D$) profiles. Panels **(b), (d),** and **(f):** Shear force versus normal force ($F_s$ vs. $F_n$) profiles, $D = 0$ nm is defined as mica-mica contact in air. Solid and open symbols represent the first and subsequent approaches, respectively. Arrows in the inset of panels (a), (c), and (e) indicate jumps in $D$ when a bilayer is removed from the contact area by hemifusion (ca. 10 to 5 nm, broken lines) or totally squeezed out (ca. 5 to 0 nm, solid line). Data for figures (b), (d), and (f) were extracted from shear force traces as in Fig. 2, whereas red broken arrows indicate increases in $F_s$ observed at hemifusion in 150 mM NaNO$_3$, while the red solid

arrows indicate 'rigid-coupling' of two surfaces at hemifusion following $Ca^{++}$ addition, where $F_s$ is higher than the shear-force detection limit of the device (185 µN) and 'rigid-coupling' occurs (see text).

L2 mixture: POPC-POPE (molar ratio 4:1), figs. 3(a), (b)

In Fig. 3a, as $F_n(D)$ increases, $D$ decreases gradually until at a critical normal load, it shifts abruptly from 9.8 ± 0.5 to 5.1 ± 1.0 nm, indicating that two PL bilayers hemifuse into one (inset to Fig. 3a). On further loading, the trapped bilayer is squeezed out of the contact area, as $D$ decreases from 4.1 ± 0.9 nm to 0.2 ± 0.4 nm, revealing that the POPC-POPE bilayer adheres only weakly to the mica surface (such total squeeze-out does not occur for the L5, L8 mixtures). On adding 2 mM calcium, the most striking change is that the normalized load ($F_n(D)/R$) that induces hemifusion increases more than 6-fold, from 80.4 ± 46.2 to 529.5 ± 207.0 mN/m, corresponding to pressures of ca. 2 – 3 MPa, comparable with those in synovial joints (see later).

The shear-force versus normal-force profiles ($F_s$ vs. $F_n$, Fig. 3b), and the shear force versus separation distance ($F_s$ vs. $D$) profiles inset to Fig. 3b show a strong increase in friction on hemifusion. Following total expulsion of the lipid, the two surfaces become rigidly coupled over the range of lateral motion applied to the top surface (so that sliding friction cannot be measured). This is expected, as the slip plane reverts from the highly-lubricated one between hydrated headgroups to a much more dissipative one between acyl tails following hemifusion.

L5 mixture: DPPC-POPC-DOPC-POPE-Egg SM (molar ratio 1:1:1:1:1), figs. 3(c), (d)

The AFM scans (Fig. 1d) together with the separation distance when two surfaces were in contact (Fig. 3c) indicate that vesicles of the 5-component mixture form flat bilayers with phase-separated domains (dimensions of order 10 – 100 nanometers) distributed irregularly on the bilayer.

These are likely composed of the saturated lipids in L5: Egg SM and DPPC, as such lipids can form gel-state domains by tight alkyl chain packing and phase-separate from the majority fluid phase composed of unsaturated PC lipids, similarly to 'rafts' on biological membranes[30]. The hybrid POPC, with one saturated and one unsaturated tail, preferentially accumulates at the interface between the gel and liquid-crystalline phases, reducing the packing incompatibility and line tension between the two phases, further stabilizing small domains in the bilayer[31,32]. After introducing 2 mM Ca(NO$_3$)$_2$, larger phase-separated domains appear on the bilayer – possibly because calcium binds laterally more strongly to Egg SM and DPPC than to POPC and DOPC[31], while the height of the patches is ca. 0.8 nm higher than the surroundings (Fig. 2f), attributed to the height difference between bilayers in gel and liquid-crystalline states (Table S1). The phase-separated regions (lighter colour) in fig. 1(e) occupy ca. 13% of the area scanned.

The normal force profiles show that the bilayer thickness reaches a 'hard-wall' at $10.4 \pm 0.8$ and $8.9 \pm 0.8$ nm, respectively, while following hemifusion, $D$ decreases to $5.1 \pm 1.1$ and $4.6 \pm 0.8$ nm respectively. Even more marked than for L2, adding calcium to the system strongly increases the critical normal load triggering hemifusion from $15.6 \pm 11.6$ to $831.0 \pm 122.3$ mN/m (fig. 3c), equivalent to contact pressures of some 4 – 5 MPa.

Prior to hemifusion, the friction is extremely low (at shear-trace noise levels) with $\mu \approx 10^{-4}$ (Fig. 3d). Immediately following hemifusion in the absence of calcium, $\mu$ increases ca. 50-fold (to $\mu \approx 5\times10^{-3}$), though at the much higher loads and pressures leading to hemifusion in the presence of 2 mM calcium, we observed a 'rigid-coupling' of two surfaces, indicating that the sliding friction exceeds the range applied shear force ($F_s > 300$ µN), corresponding to $\mu \gtrsim 3\times10^{-2}$.

*L8 mixture*: DPPC:POPC:DOPC:POPE:Egg SM:O-LPC:CHOL:DPPA (molar ratio 1:1:1:1:1:1:1:0.1), figs. 3(e), (f)

The L8 mixture consists of representatives of essentially all the major PL groups identified in synovial joints, at roughly their proportions as measured on articular cartilage surfaces[22] (fig. S2). It includes, in addition to the lipids in L5, also lyso-PC (O-LPC), cholesterol (CHOL), and a negatively charged PL (DPPA, 1.2 mol%). Lyso-PC, with a conical configuration (positive curvature), is more likely to assemble in the outer leaflet of vesicles, and inhibits membrane fusion[33]. CHOL is ubiquitous in PL cell membranes, and it is also found in synovial fluid[34]. CHOL has an affinity to different PLs in the decreasing order SM > PC > PE, and preferentially interacts with saturated PCs over unsaturated ones; it has been shown both in experiments and in molecular dynamics simulations to promote lateral segregation into phases rich in either saturated or in unsaturated lipids[35,36]. Adding CHOL to single-component PCs promotes bilayer hemifusion[17]. The gel-state negatively-charged PL, DPPA, not only brings negative charges to the gel-state patches but also shows particularly strong interaction with cations, particularly multi-valent cations, such as $Ca^{++}$[37]. We emphasize however, as considered further in the Discussion, that although all these lipid types are found in joints, it is not known whether they comprise part of the lubricating boundary layers on articulating cartilage *in vivo*.

DLS measurements (Table 1) show that the L8 vesicles are monodispersed and somewhat negatively charged. Adding calcium induces aggregation of liposomes, revealed by multiple peaks with larger sizes and a higher polydispersity (Table 1 and Fig. S3), as well as a slight reduction in the ζ-potential. On adsorption to mica the liposomes rupture and form a planar layer with round phase-separated domains with diameter in the range of tens of nanometers (Fig. 2g). These may be attributed to the promotion by the cholesterol of phase separation of saturated from the unsaturated lipids in L8[35,36]. After adding 2 mM Ca(NO$_3$)$_2$, larger phase-separated gel-state patches are observed, and the height of these domains is ca. 1-2 nm higher than the surroundings (Figs. 2h and 2i). This may arise from calcium binding strongly to adjacent anionic PLs, thus promoting a tighter packing of alkyl chain in the membrane and a thicker hydrophobic region[38].

The normal force profiles in Fig. 3e show that, as for L2 and L5, hemifusion of the compressed L8 bilayers is strongly suppressed and moved to higher pressures by the presence of calcium. Thus, hemifusion is observed at critical normal loads 48.5 ± 34.6 (in the range of 19.1 – 157.1) mN/m and 384.8 ± 111.5 (in the range of 241.3 – 610.1) mN/m before and after adding 2 mM calcium salt, while $D$ decreases from 8.1 ± 0.8 and 7.0 ± 1.1 nm to 2.6 ± 0.5 and 3.1 ± 0.4 nm, respectively.

The measured friction behaviour (Fig. 3f) is similar to the other mixtures, with friction increasing strongly upon hemifusion so that the two surfaces become rigidly coupled on lateral motion of the top one (corresponding in this case to $\mu \geq$ ca. 0.07), though most of the critical loads (and thus contact pressures, see below) at hemifusion are lower than for L2 and L5,.

**Discussion and conclusions**

The main findings of this work are as follows: At low loads (low contact pressures) all three synovial lipid mixtures display excellent boundary lubrication properties ($\mu \approx 10^{-4}$ or lower). As the normal loads (and thus the contact pressures) increase beyond some critical value, the layers undergo hemifusion, at which point the friction rises abruptly. Addition of Ca$^{++}$ to physiological concentrations (2 mM) greatly increases the magnitude of these critical loads, so that hemifusion occurs at higher contact pressures, comparable with those at articular cartilage in synovial joints during free walking (around 5 MPa)[39].

A closer examination however reveals, remarkably, that using the 5-component lipid mixture L5 leads to a lubricating layer which is more robust to hemifusion under loading and shear than either the 2-component L2 or the 8-component L8 mixtures, in the sense of sustaining very significantly higher pressures before hemifusion and a sharp rise in the friction occur. This result, shown in fig. 4 and considered further below, indicates lubrication synergy in suitable mixtures of lipids. It is both unexpected and strongly suggestive, shedding light on the possible origin of the proliferation of lipid-types in joints from the view-point of cartilage lubrication. It is instructive first to consider the effect of the presence of calcium on the interactions between the lipid- coated surfaces.

**Effect of Calcium**

In the context of lubrication, one might have expected that the anionic $Ca^{++}$ ions would bridge the opposing lipid layers through adhesive dipole-charge or charge-charge interactions. Indeed, the aggregation induced by $Ca^{++}$ ions for vesicles of the L8 mixture (Table 1) indicates that such bridging occurs between the negatively-charged DPPA headgroups. Adhesive bridging might be expected to increase frictional dissipation as the surfaces slide, due to hysteretic bond breakage and reformation. In contrast to this, our results reveal the opposite, counterintuitive effect: that adding physiological-level concentrations of calcium actually improves the lubrication, for all three mixtures studied, by suppressing the hemifusion of opposing bilayers, and hence the increase in friction, up to much higher loads than is the case in the absence of such ions.

We may attribute this suppression of hemifusion to the fact that the added calcium increases the *intra*-layer cohesion by bridging adjacent zwitterionic phosphocholine, leading also to a higher areal density of the lipids[37]. This increased cohesion implies that larger loads/pressures are then needed to induce hemifusion. Our findings contrast with a previous report that physiological level calcium promotes hemifusion of supported PL membranes[40]. The difference arises because in the previous study an asymmetric bilayer was used, where the lower leaflet rigidly anchored the outer, negatively-charged, mixed-lipid leaflet to the substrate, and the $Ca^{++}$ ions bridged the two bilayers strongly to induce hemifusion. This contrasts with the present study where calcium ions strengthened the much weaker bilayer/substrate interaction, while their intra-layer interactions within the essentially-neutral upper leaflets, enhanced by $Ca^{++}$ as described above, rendered them more robust against hemifusion.

**Lipid synergy in articular cartilage lubrication**

The central question that this study addresses, as posed in the Introduction, is whether the presence of many different lipid types in synovial joints could, apart from any other biological roles that they play, lead to synergy in the boundary lubrication of articular cartilage. In other words, could the presence of a particular mix of lipids in the boundary-layer coating the cartilage, composed of those lipids present in the joint, result in optimal lubrication, in the sense of lowest friction up to the highest (physiological) contact pressures? Such synergy is indeed directly indicated by our observations on lubrication by boundary layers of the L2, L5, and L8 lipid mixtures as summarized in Fig. 4 showing the critical loads at hemifusion with boundary layers of these three different mixtures.

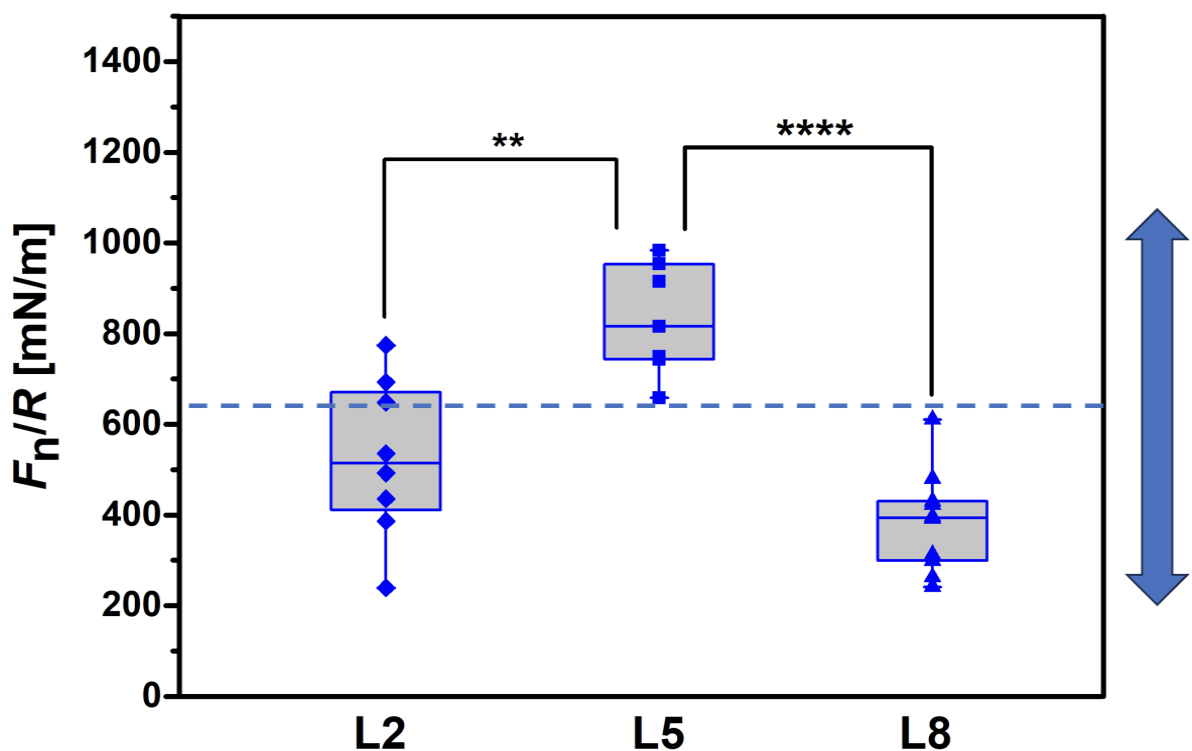

**Figure 4.** Normalized forces at the critical loads for hemifusion in the presence of physiological-level calcium for the three systems. Data are presented as box plots, whereas each box shows the interquartile range and the line inside represents the median. ** $p \leq 0.01$, **** $p \leq 0.0001$. The blue double-arrow shows the corresponding range of maximal contact pressures at the articular cartilage surface directly measured with pressure transducers in the human hip joint during free walking[39], while the broken blue line is its mean value. Statistical analysis (unpaired *t* test) was performed using GraphPad Prism.

As seen clearly in Fig. 4, the boundary layers composed of the L5 mixture, comprising DPPC, POPC, DOPC, POPE, and Egg SM in equimolar concentrations, can provide low friction ($\mu \lesssim 10^{-4}$, fig. 3(d)) prior to hemifusion up to mean critical loads that are some 50% and 100% higher than with boundary layers composed of the L2 or L8 mixtures respectively. Thus a clear synergy arising from the multiplicity of lipids is achieved in this case. This superior lubrication ability of L5 arises by adding to the L2 mixture (POPC and POPE) the additional lipids DPPC, DOPC, and Egg SM. The mere addition of more different lipid types is not in itself the origin of the better lubrication, as seen for L8 where the components of L5 are further augmented by other lipid types found in synovial joints, but where the lubrication is very significantly inferior to that of L5 as seen in Fig. 5. Our point is not that L5 represents the actual composition of lubricating boundary layers on articular cartilage, nor even that it is the best possible combination of the lipids used in the present study (Fig. S1 and Table S1) for forming lubricating boundary layers. Rather, the fact that a particular combination of lipids that are present in synovial joints, as in L5, provides superior lubrication – in the sense of a lubricating layer that is more robust to pressure and shear - than other combinations, is an unequivocal proof-of-concept that the proliferation of lipids in joints may indeed provide lubrication synergy. One other feature of fig. 4 may be emphasized: the mean maximal pressure on human articular cartilage during walking, indicated by the broken line in fig. 4, is very much in the range where lubrication breakdown due to hemifusion occurs in the mixtures used in this study. But while the mean L2 and L8 breakdown values are below this mean value, that of L5 is above it; simplistically, that suggests that a lubricating boundary layer of L5 lipids would lubricate human joints – and avoid cartilage-wear leading to osteoarthritis – much more efficiently than layers of either L2 or L8.

Can we determine what an optimally-lubricating lipid mixture might consist of? In principle one may measure directly the behaviour of many different lipid combinations to identify the best one, but it is clearly impractical to measure more than a tiny fraction of all possible combinations of the 100+ lipids in synovial joints (thus there are ca. $10^{10}$ ways of choosing, say, a 5-lipid combination from 100 different lipids, and even a combination of 5 out of the 50 most likely lipids presents over $2.10^6$ possibilities). Alternatively, one may try to predict the lubrication properties of a lipid mixture by considering the attributes of its component lipids. For example, saturated PC lipids (such as DPPC in L5) have highly hydrated headgroups promoting low friction through the hydration lubrication mechanism[14], and are in the mechanically-robust gel-phase. On the other hand , they may be slow to heal once damaged[11,19], and their bilayers are associated with a friction coefficient that is higher than that of unsaturated lipids (e.g. POPC[11]). Such 'pros' and 'cons' can be considered also for other lipids (Table S2). However, it is far from clear how lubrication by a mixture of lipids relates to the sum of its components; for example, phase separation and inter-lipid interactions may well modulate the bilayer behaviour in a non-additive manner.

**Stalk formation and hemifusion**

In view of the complexity of property-additivity and the consequent difficulty of heuristically-identifying optimal combinations based just on their single-lipid properties (e.g. Table S2), we propose a different approach to get insight into lipid mixtures that may have better boundary lubrication properties. This relies on molecular dynamics (MD) to probe both frictional properties and hemifusion of compressed, sliding bilayers consisting of different lipid mixtures. Very recently we used all-atom MD to evaluate friction between two (single component) POPC lipid bilayers sliding past each other[41,42], showing good quantitative agreement with the SFB-measured friction, and this approach can readily be extended also to friction between bilayers consisting of lipid mixtures as in the present study.

For probing the likelihood of *hemifusion* of bilayers, as seen in fig. 3 for the mixtures used in this study, which is the limiting factor for their efficient lubrication, a different MD approach may be used. A well-known indicator of impending hemifusion is the formation of a *stalk structure* between interacting membranes (i.e. lipid bilayers)[43-45], as schematically shown in fig. 5A below. The development of such a stalk may be monitored graphically, and its likelihood may be gauged by evaluating the potential of mean force (PMF) $\Delta G_{PMF}$ at different values of the reaction coordinate $\xi_{ch}$ corresponding to different stages of the stalk formation[43]. Higher values of $\Delta G_{PMF}$ correspond to a lower likelihood of stalk formation and therefore of hemifusion.

As proof-of-concept that this may yield insight into the lubrication-robustness (i.e. resistance to hemifusion) of lipid mixture bilayers, we carried out such MD calculations of $\Delta G_{PMF}$ for the lipid mixtures used in this study (L2, L5 and L8), using the same MD protocols as in ref.[43], with hydration levels $n_w$ = 5 or 12 water molecules/lipid (see SI for detailed method). These respectively represent the critical hydration level for hemifusion and the full hydration level for unperturbed bilayers[43]. For comparison with experiment, we use the lower bound ($n_w$ = 5 water molecules/lipid), since hemifusion occurs under strong compression in the SFB, where the number of water molecules/headgroup reduces locally due to pressurized deformation across the contact area. In the case of the L5 mixture, we calculate $\Delta G_{PMF}$ for the as-prepared composition, but we note that the AFM micrograph (fig. 1(e)) indicates some phase separation of the gel-phase lipids, as noted earlier. Since the majority liquid-phase regions of L5 are more prone to hemifusion, we calculate $\Delta G_{PMF}$ also for these. Estimating for L5 the extent of phase separation from fig. 1(e) as 13% (of the total of 40% gel phase lipids DPPC and Egg SM), we examine stalk formation between a mixture of the majority liquid-phase lipids together with the residual 27% of DPPC and Egg SM, for two limiting cases (fig. 5C, D). Stalk formation is more likely between such liquid-phase-lipids enriched mixtures, and since hemifusion initiated at such domains would lead to hemifusion of the entire bilayer, they constitute (and are designated in fig. 5) as the 'weak link' for stalk formation and hemifusion. For the case of the L8 mixture, the situation is more complicated, since that mixture contains CHOL which is known to cause phase separation in bilayers of mixtures of saturated and unsaturated lipids[35,36], implying – for L8 - phase-separated domains of POPC and DOPC (enriched in CHOL). Interacting bilayers of CHOL-rich DOPC have earlier been shown to have lower $\Delta G_{PMF}$ values for stalk formation than the CHOL-free DOPC bilayers[43], and as for the L5 domains above, they would constitute the 'weak link' for stalk formation and hemifusion for L8. We therefore also examine stalk formation between DOPC/CHOL and between DOPC/POPE/CHOL layers.

The results on the likelihood of stalk formation for the three lipid mixtures (including the 'weak-links' for L5 and L8) are presented in fig. 5C, D.

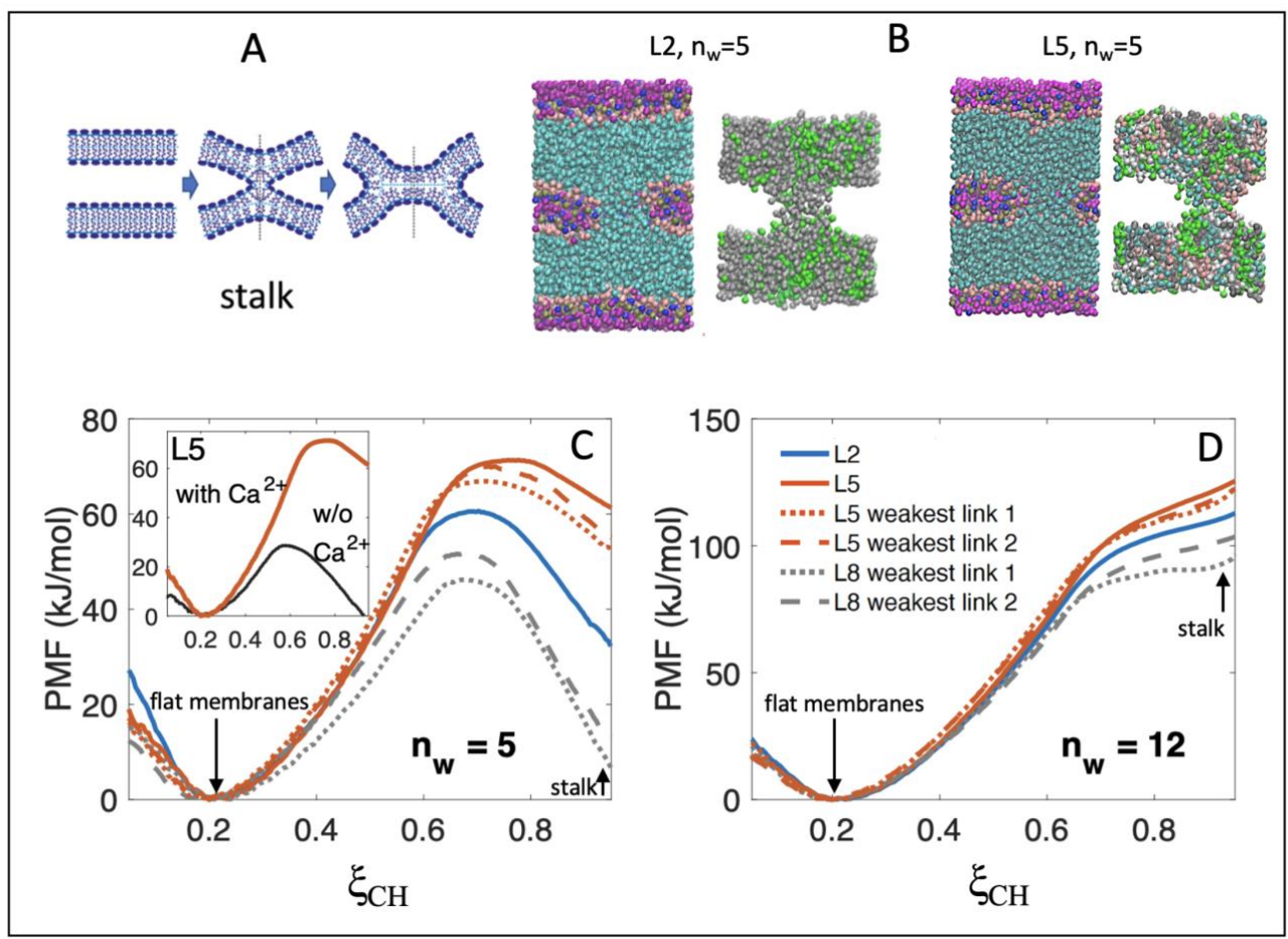

**Figure 5**. MD simulations of stalk formation in lipid mixtures mixtures. (A) shows schematically how stalk formation leads to hemifusion and expulsion of a bilayer. (B) illustrates graphically the stalk formation in the L2 and L5 mixtures at hydration level $n_w = 5$ water molecules per lipid. L2, <u>Left</u>: magenta: water; cyan: acyl tails; gold: phosphate; blue: choline headgroup; pink: glycerol moieties. <u>Right</u> (acyl tails only): grey: POPC; green POPE. L5, <u>Left</u>: magenta: water; cyan: acyl tails; gold: phosphate; Blue: choline headgroup; pink: glycerol moieties. <u>Right</u> (acyl tails only):

grey: SM (egg sphingomyeline); pink: POPC; green DPPC; white: DOPC; cyan: POPE. (C) and (D): The potential of mean force (PMF) plotted as a function of the reaction coordinate $\xi_{ch}$ (extent of stalk formation), at $n_w = 5$ and $n_w = 12$ hydration levels, respectively, in presence of $Ca^{++}$, for the following: L2; L5 (as prepared); 'weakest links' 1 & 2 for L5; and weakest links 1 & 2 for L8 (see text), colour-coded as in legend in panel (D). For all PMF profiles, the estimated error is < 3 kJ/mol. For L5, POPC: DOPC: POPE: DPPC: DPSM, weakest-link 1 has molar ratios 1 :1 :1:1 :0.27, and weakest link 2 is 1 :1 :1 :0.6 :0.67. For L8, weakest link 1 is DOPC:CHOL in molar ratios 1:0.16, and weakest link 2 is DOPC: POPE: CHOL in ratios 1 :1 :0.3 (see text). Inset to (C): the effect on PMF for L5 of adding $Ca^{++}$.

Fig. 5D shows that the magnitude of $\Delta G_{PMF}$ for the as-prepared L5 mixture is higher than for the L2 mixture, and remains significantly higher even when phase-separation, as indicated in fig. 1(e), is taken into account as noted above ('weakest-link' domains), i.e. this result predicts clearly that L5 would better resist hemifusion than L2. Likewise, for CHOL-induced phase-separated regions in L8, the magnitude of $\Delta G_{PMF}$ is significantly lower than for L2, predicting more likely stalk formation, and thus hemifusion, for the 8-component mixture relative to the 2-component one. This ranking of hemifusion likelihood holds for both $n_w = 5$ and $n_w = 12$ cases, though it is more marked in the former where stalk formation is an energetically favoured state. These prediction of the MD calculations, i.e. that the likelihood of hemifusion decreases as L8 > L2 > L5, so that the critical pressure to induce hemifusion increases as L8 < L2 < L5, are exactly what is seen in the experimental SFB results, fig. 4. This behaviour could not have been predicted a priori based simply on mixing of the qualitative properties of the single lipid bilayers (e.g. Table S2). This proof-of-concept demonstration thus strongly supports the notion that MD may be a powerful tool in gaining insight into nature's multi-lipid synergy in boundary lubrication of cartilage. We have also examined the effect on the PMF of adding $Ca^{++}$ to the lipid mixtures, as in the inset to fig. 5D: the presence of $Ca^{++}$ leads to a very marked increase in the PMF, i.e. stalk formation and thus hemifusion is suppressed, precisely as seen in the SFB experiments. This effect too could not have been predicted a priori by heuristic considerations, indeed it is somewhat counterintuitive as earlier discussed. It is finally of interest that stalk formation as a precursor to hemifusion has to date been considered mostly in the biochemical context of cell membrane interactions or of vesicle-cell fusion; here it plays a central role in lipid-mediated lubrication, with relevance to cartilage lubrication and its connection to joint homeostasis.

To conclude: we have shown that mixtures of lipids present in synovial joints may form boundary layers that possess excellent lubrication properties – comparable with those of healthy joints up to physiological pressures – while at the same time they may combine desirable features of their different components. By revealing that particular combinations of these lipids (e.g. L5) can be significantly superior as lubricating layers compared to other combinations with either more (L8) or fewer (L2) lipid components, we unambiguously demonstrate the possibility of multi-lipid synergy. While we don't claim to have identified the optimal composition of such a layer, it is nonetheless clear from our results on a limited sample of synovial joint lipids that such an optimal composition is possible. This may advance our understanding of the proliferation of different lipid types in healthy synovial joints: purely from a lubrication point of view, essential for joint homeostasis, such a proliferation, enabling optimal lubrication, is clearly beneficial. Importantly, we were able to show, in a proof-of-concept demonstration, that molecular dynamics simulations are able to predict accurately the relative robustness against hemifusion of the different lipid mixtures (specifically, that L5 is less likely to hemifuse than L2, which in turn is less likely to hemifuse than L8, which indeed were key observations). In the light of recent suggestions[2] that intra-articularly (IA) injected liposomes may serve to augment or repair the body's natural biolubrication mechanisms at the articular cartilage surface, our results may also point how to identify and implement optimal liposome compositions for such IA administration.

## Materials and Methods

### Materials

Eight different lipids as detailed below representing the main types in human synovial joints were used in this study: Egg sphingomyelin (Egg SM, whose fatty acids distribution is 86% 16:0, 6% 18:0, 3% 22:0, 3% 24:1, and 2% unknown) and 1-oleoyl-2-hydroxy-*sn*-glycero-3-phosphocholine (O-LPC, purity > 99% LPC, may contain up to 10% of the 2-LPC isomer), and 1-palmitoyl-2-oleoyl-glycero-3-phosphoethanolamine (POPE) were purchased from Avanti Polar Lipids, Inc. (Alabama, USA). DPPC, POPC, 1,2-dioleoyl-*sn*-glycero-3-phosphocholine (DOPC), and 1,2-palmitoyl-phosphatidic acid (sodium salt) (DPPA) were obtained from Lipoid GmbH (Ludwigshafen, Germany). Cholesterol ($\geq$ 99%), sodium nitrate ($NaNO_3$) 99.99 Suprapur®, and calcium nitrate tetrahydrate ($Ca(NO_3)_2 \cdot 4H_2O$) 99.95 Suprapur® were purchased from Merck KGaA (Darmstadt, Germany). Chloroform (analytical reagent grade) and methanol (HPLC grade) were purchased from Bio-Lab Ltd. (Jerusalem, Israel). Conductivity water (resistivity 18.20 M$\Omega$·cm, total organic carbon content $\leq$ 2 ppb) was obtained with a Thermo Scientific™ Barnstead™ water purification system. The chemical structures and physiochemical properties of these 8 lipids (POPC, POPE, DPPC, DOPC, Egg SM, O-LPC, Chol and DPPA) are shown in Fig. S1 and Table S1.

### Preparation of SUVs

The preparation of SUVs was carried out using a hydration – extrusion method. Lipids were mixed by dissolving separately in a solvent of chloroform and methanol (2:1, v:v) except for POPE, which was dissolved in chloroform. After that, appropriate amounts of different PL solutions were thoroughly mixed in a glass vial. Organic solvents were then removed by purging nitrogen gas for at least one day and followed by lyophilization overnight. The SUVs were prepared by a thin-film hydration followed by extrusion method. The procedures were similar to those described previously[19] except for a few modifications: a 150 mM $NaNO_3$ aqueous solution was used as dispersant and temperature was controlled at ca. 10 °C higher than the highest phase transition temperature of individual PL components for both sonication bath and extruder jacket. The prepared SUV dispersions were cooled down to room temperature and kept at 4 °C before use.

**Size distribution, polydispersity index (PDI), and zeta potential**

The prepared SUVs were measured on a Malvern Zetasizer Nano ZS instrument. Size analysis were performed with a backscattering angle at 173°, and the values were determined according to the Stokes-Einstein equation. Liposomes with a total lipid concentration of 0.3 mM prepared in 150 mM $NaNO_3$ or after adding 2 mM $Ca(NO_3)_2$ were directly used for size distribution measurements, and were diluted by ten-fold with water for zeta potential measurements.

**Atomic force microscopy (AFM)**

The AFM was used to probe the morphology of SUVs on mica. An Asylum MPF-3D atomic force microscope and Bruker's SNL-10 probes were used. All the scans were performed under aqueous environment in tapping mode. Samples were prepared by introducing 0.3 mM SUVs in 150 mM $NaNO_3$ or in 150 mM $NaNO_3$ with 2 mM $Ca(NO_3)_2$ to a petri dish with a freshly cleaved mica facet glued on the bottom. After incubating for more than 4 hours, scans were performed under the same aqueous solution.

**Surface force balance (SFB)**

The normal ($F_n$) and shear forces ($F_s$) at different surface separations ($D$) between two back-silvered mica surfaces in a crossed-cylinder configuration with and SFB. A schematic of the SFB setup is shown in fig. 2a. Its details and the experimental procedures have been elaborated previously[46]. Briefly, the wavelengths of multiple-beam interference fringes of equal chromatic order (FECO) transmitted between the surfaces were monitored to determine $D$; $F_n$ was calculated according to the relative difference between the displacement of the normal spring (spring constant $k_n$) and the applied displacement; and $F_s$ was calculated by applying fast Fourier transform (FFT) to the recorded lateral displacement of the shear spring (spring constant $k_s$), which were monitored by an air-gap capacitor as responses to the lateral back-and-forth movements applied to the upper lens through the piezoelectric tube (PZT). The friction coefficient $\mu$ was calculated by the ratio of shear force $F_s$ required to slide the surfaces and the corresponding normal load $F_n$ compressing them, that is, $\mu = F_s/F_n$.

Before adding calcium, the two surfaces were non-adhesive and it was difficult to measure the flattening of the fringes. Thus, normal pressure at hemifusion $P_{hemi}$ was estimated via the Hertzian model, $P_{hemi} = F_{n,hemi}/(\pi(F_{n,hemi} \cdot R/K)^{2/3})$, where $K = 5 \times 10^{-9}$ N/m$^2$ was adopted as the effective elastic modulus of the mica/glue combination[14]. After adding calcium, the normal pressure $P_{hemi}$ applied to the contact area at hemifusion was calculated directly by the measurement of normal force $F_{n,hemi}$ and measured fringe flattening radius $a$ at hemifusion, $P_{hemi} = F_{n,hemi}/A = F_{n,hemi}/(\pi a^2)$. We note that while the magnitude of $F_{n,hemi}$ is accurately known from the extent of measured spring-bending, the contact area *a* is less accurately measured from the fringe-tip flattening, particularly when it is not large, so that a difference in *P* values of up to ± 30% or sometimes more under the same conditions may be obtained by these two methods[14].

At the beginning of the measurements, mica-mica contact in air corresponding to $D = 0$ nm was measured. After that, the boat was filled with 150 mM $NaNO_3$ solution to which was then added 0.8 mL of 6 mM mixed PLs-SUV dispersion prepared in the same salt solution to reach a final lipid concentration at ca. 0.3 mM, in the range of values reported in healthy synovial fluid[24]. Measurements were carried out after incubating for more than four hours to reach an equilibrium. After force profile measurements, $Ca(NO_3)_2$ aqueous solution was introduced to the boat, reaching a final concentration of 2 mM, which is in the mid-range of those identified in human synovial fluid[47], and force profiles were measured again. All the data presented below were collected from at least two independent experiments and several contact points in each experiment.

**MD simulations of the potential of mean force (PMF) calculation of the stalk formation.**
The PMF calculation was performed with the protocol described in Ref.[43] and carried out with the modified GROMACS version published by Ref. [43] in GitLab repository. It implemented a stalk formation reaction coordinate $\xi_{CH}$ which describes the hydrophobic connectivity between the two bilayer membranes and allows a stalk to form artificially by applying a harmonic restrain along the reaction coordinate. To acquire a single PMF, multiple steps of GROMACS simulations, manipulation of structure files, and indexing of groups within the structure were carried out, which were all automized using in-house Bash scripts. The steps are briefly described below.

The structures of the single bilayers were generated from CHARM-GUI interface using the Martini coarse-grained force field version 2.2 [48-50]. Each single membrane contained 64 lipids/monolayer, sufficient to exclude the size effect for the L2 and L5 mixtures which are free of cholesterol. For the cholesterol-containing L8 mixture, the energy barrier to stalk formation $\Delta G_{PMF}$ may be overestimated but does not change our conclusion that the weakest links of the L8 mixture is the least resistive to hemifusion in all systems tested (see main text) [43]. 0.5 mole $CaCl_2$ / mole lipid was added to the system with parameters taken from the Martini 2.0 force field. $Ca^{2+}$ and $Cl^-$ ions are represented as Lennard-Jones particles with parameters calibrated inclusive of the hydration-shell. Note that due to the limitation of the Martini description of divalent ions, the results are not to be considered quantitatively, but for our goal of ranking the likelihood of stalk formation of membranes treated with the same salt solvent, it is sufficiently representative. The energy of the single membrane was then minimized. Excess water was then removed from the system such that the hydration level of the outer leaflet is 36 water molecules/lipid, and the hydration level of the inner leaflet is 5 or 12 water molecules/lipid. The single membrane was then replicated, flipped vertically, translated and stacked to obtain a symmetric double-bilayer system. By assuming a hydration shell of 7 and 12 water molecules / $Ca^{2+}$ respectively for the dehydrated and fully hydrated cases based on NMR studies [51], we estimated that the $Ca^{2+}$ concentration in the inner compartment of the systems of the two hydration levels $n_w=5$ and $n_w=12$ are respectively 1.3 M and 2.2 M. This is consistent with previous second harmonic microscopy observations, that the ion concentration in the lipid headgroup region is 1-3 M [52,53].

    The double-bilayer system then underwent energy minimization and equilibration for 25 ns until the box dimensions and energy were stabilized. The equilibrated system was processed following the exact steps and GROMACS parameters and restrains published by Ref. [43]. First, the carbon tails of the systems were pulled using the harmonic potential ($k = 3000$ kJ/mol) into the inner inter-membrane space to artificially induce the stalk structure within 200 ns. For the umbrella

sampling, 19 initial structures were extracted from the trajectories evenly along the reaction coordinate $\xi_{CH}$. PMFs were calculated using umbrella sampling with these initial frames using the same harmonic restrain at the corresponding $\xi_{CH}$ values, with each run persists 200 ns at a time step of 18 fs. The PMFs were then calculated using the gmx wham function of GROMACS 2023.2. The error is estimated following the method proposed in Ref. [43], where separated PMFs were generated using pieces of the umbrella sampling simulations, respectively at 50-100 ns, 100-150 ns, and 150-200 ns intervals. The error is estimated as the standard deviation between these PMF curves. The average error along a given PMF curve is 1.0-1.6 KJ/mol while the maximum error is 2.1-3.3 KJ/mol.

**Conflict of interest**

The authors declare no competing interest.

**Acknowledgement**

We thank the European Research Council (Advanced Grant CartiLube 743016), the McCutchen Foundation, the Israel Science Foundation – National Natural Science Foundation of China joint research program (Grant 3618/21), the Israel Ministry of Science and Technology (Grant 3-15716), and the Israel Science Foundation (Grant 1229/20) for financial support. This work was made possible partly through the historic generosity of the Perlman family.